\documentclass[sigconf,natbib=false]{acmart}

\acmConference[ICPC 2024]{46th International Conference on Program Comprehension}{April 2024}{Lisbon, Portugal}

\newcommand{\squishlist}{
 \begin{list}{$\bullet$}
   { \setlength{\itemsep}{0pt}
     \setlength{\parsep}{0pt}
     \setlength{\topsep}{0pt}
     \setlength{\partopsep}{0pt}
     \setlength{\leftmargin}{2.5em}
     \setlength{\labelwidth}{1.5em}
     \setlength{\labelsep}{0.5em} } }

\newcommand{\squishend}{
  \end{list}  }

\usepackage{colortbl}
\usepackage{amsmath}

\AtBeginDocument{%
  }

\copyrightyear{2024}
\acmYear{2024}
\setcopyright{acmlicensed}\acmConference[ICPC '24]{32nd IEEE/ACM International Conference on Program Comprehension}{April 15--16, 2024}{Lisbon, Portugal}
\acmBooktitle{32nd IEEE/ACM International Conference on Program Comprehension (ICPC '24), April 15--16, 2024, Lisbon, Portugal}
\acmDOI{10.1145/3643916.3644434}
\acmISBN{979-8-4007-0586-1/24/04}

\usepackage{tabularx}

\begin{document}


%
\title{Do Machines and Humans Focus on Similar Code? Exploring Explainability of Large Language Models in Code Summarization}




\author{Jiliang Li}
\affiliation{%
 \institution{Vanderbilt University}
 \city{Nashville}
 \state{Tennessee}
 \country{USA}}
\email{jiliang.li@vanderbilt.edu}

\author{Yifan Zhang}
\affiliation{%
 \institution{Vanderbilt University}
 \city{Nashville}
 \state{Tennessee}
 \country{USA}}
\email{yifan.zhang.2@vanderbilt.edu}

\author{Zachary Karas}
\affiliation{%
 \institution{Vanderbilt University}
 \city{Nashville}
 \state{Tennessee}
 \country{USA}}
\email{z.karas@vanderbilt.edu}

\author{Collin McMillan}
\affiliation{%
 \institution{University of Notre Dame}
 \city{Notre Dame}
 \state{Indiana}
 \country{USA}}
\email{cmc@nd.edu}

\author{Kevin Leach}
\affiliation{%
 \institution{Vanderbilt University}
 \city{Nashville}
 \state{Tennessee}
 \country{USA}}
\email{kevin.leach@vanderbilt.edu}

\author{Yu Huang}
\affiliation{%
 \institution{Vanderbilt University}
 \city{Nashville} 
 \state{Tennessee}
 \country{USA}}
\email{yu.huang@vanderbilt.edu}


\begin{abstract}

Recent language models have demonstrated proficiency in summarizing source code. 
However, as in many other domains of machine learning, language models of code lack sufficient explainability --- informally, we lack a formulaic or intuitive understanding of what and how models learn from code.
Explainability of language models can be partially provided if, as the models learn to produce higher-quality code summaries, they also align in deeming the same code parts important as those identified by human programmers.
In this paper, we report negative results from our investigation of explainability of language models in code summarization through the lens of human comprehension.
We measure human focus on code using eye-tracking metrics such as fixation counts and duration in code summarization tasks. 
To approximate language model focus, we employ a state-of-the-art model-agnostic, black-box, 
perturbation-based approach, SHAP (SHapley Additive exPlanations), to identify which code tokens influence that generation of summaries.
Using these settings, we find no statistically significant relationship between language models' focus and human programmers' attention.
Furthermore, alignment between model and human foci in this setting does not seem to dictate the quality of the LLM-generated summaries.
Our study highlights an inability to align human focus with SHAP-based model focus measures.
This result calls for future investigation of multiple open questions for explainable language models for code summarization and software engineering tasks in general, including the training mechanisms of language models for code, whether there is an alignment between human and model attention on code, whether human attention can improve the development of language models, and what other model focus measures are appropriate for improving explainability.  

\end{abstract}


%



\maketitle

\section{Introduction}

Recent language models for code have shown promising performance on several code-related tasks~\cite{xu2022systematic}.
Among these tasks is neural code summarization, where a language model generates a short natural language summary describing a given code snippet. This is often an indicative task demonstrating a model's ability to comprehend code.
Currently, the majority of assessments for how well a language model understands code 
directly measures the quality of code summaries generated by the models, and compares them with human-written summaries~\cite{xu2022systematic}.
Comparatively little is known about why and how the language models reason about code to generate such summaries.
Similar to many other downstream domains of machine learning in software engineering, understanding and explaining how and why language models for code work (or fail) 
is critical to improving model architecture, reducing bias, and preventing undesirable model behavior.

Human programmers typically achieve a strong understanding of code. Thus, proficient language models might be explained if they focus on the same parts of code that humans would~\cite{paltenghi2021thinking}.
Eye-tracking studies have been conducted to analyze programmers' visual patterns while reading code~\cite{abid2019using, rodeghero2015empirical}. 
Specifically, 
the duration and frequency of a programmer's eye gaze on a part of code in a spatially-stable manner, referred to as \textit{fixation duration} and \textit{fixation count} respectively, are indicative of cognitive load~\cite{sharafi2015eye}. 
Thus, these measures of eye-tracking can indicate the parts of code on which human programmers focus.
In contrast, 
there is a lack of consensus on how to measure a language model's reasoning about code (see Section~\ref{related:attention}). Most existing works extract the self-attention layers in language models for code to measure the model attention~\cite{paltenghi2021thinking, paltenghi2022extracting, huber2023look}. Such methods require direct access to the internal layers of a language model, limiting the possibility to investigate interpretability of many state-of-the-art proprietary models (e.g., ChatGPT).

In this paper, to investigate how proprietary language models reason about code, we employ a state-of-the-art perturbation-based method, SHAP~\cite{lundberg2017unified} (SHapley Additive exPlanations), that treats each language model as a black-box function. With SHAP, we analyze the feature attribution (i.e., which parts of code are deemed important by the model) in six different state-of-the-art language models for code. 
We use a set of Java methods to task both the language models and human programmers with writing code summaries. The feature attribution in the language models, measured by SHAP, is then compared with human developers' focus, collected from eye-tracking.
We hypothesize that sufficiently large models may learn to focus on parts of code similarly to humans. 
If validated, language model behavior can thus be described in terms of human behavior, ultimately helping to explain and improve language models.
However, we find that explainability cannot be provided through this lens and find no statistically significant evidence suggesting the hypothesized alignment. 
Furthermore, we did not find that language models' focus exhibits a statistically significant correlation with human focus in general. 
For future research that aims to explore the explainability of language models for code summarization, especially for those leveraging human attention, our findings might suggest the following: 
(1) though widely used in AI, SHAP may not be an optimal method to investigate where language models focus during code summarization, or alternatively,
(2) a misalignment between language models and human developers in reasoning about code
may provide insights for improving AI models for code summarization.

\section{Background and Related Work}

\subsection{Neural Models for Code Summarization}\label{related:model}
Advancements in deep learning have enabled machine learning models to generate summaries for source code.
Among the state-of-the-art models, NeuralCodeSum (NCS) first introduced the use of Transformers in neural code summarization~\cite{ahmad2020transformer}.
With the rise of large language models (LLMs), ServiceNow and HuggingFace released a 15.5B parameter LLM for code, StarCoder~\cite{li2023starcoder}, and Meta released a 7B parameter LLM, Code LLama~\cite{roziere2023code}, both of which can serve to summarize code.
Although not inherently an LLM for code, GPT3.5~\cite{openai2023chatgpt} and GPT4~\cite{openai2023gpt} are also capable of code summarization.
In this paper, we investigate how all the aforementioned models reason about code when tasked to generate code summaries.

\subsection{Interpretability of Language Models}\label{related:attention}
Existing works on interpretable language models generally seek to investigate the relative importance of each input token for model performance~\cite{vaswani2017attention, hooker2018evaluating, molnar2020interpretable}.
Such works can be commonly categorized into two types: white-box vs. black-box.
White-box approaches require access to a language model's internal layers~\cite{shrikumar2017learning, sundararajan2017axiomatic}, often directly investigating the self-attention scores in Transformer-based models~\cite{galassi2020attention, zhang2022does, zeng2022extensive}.
However, Transformer-based models' inherent complexity has led to a lack of consensus on how to aggregate attention weights~\cite{wang2021wheacha, zhang2022diet, zhang2022does}.
For the general research community, white-box approaches preclude proprietary models (e.g., ChatGPT).

In contrast, state-of-the-art black-box approaches like SHAP~\cite{lundberg2017unified} (SHapley Additive exPlanations) apply game-theoretic principles to assess the impact of input variations on a model's output.
SHAP evaluates the effects of different combinations of input features --- such as tokens in a text sequence --- by observing how their presence or absence (simulated by token masking) alters the model's prediction from an expected result.
This process helps to ascertain the relative contribution of each feature to the output, allowing for an analysis of the model without requiring access to its internal architecture~\cite{liu2023reliability, kou2023model}.
In this paper, to investigate proprietary models, we employ SHAP to measure where language models focus on code.

\subsection{Comparing Human vs. Machine Attention}
Previous papers have examined the alignment between human and model attention in code comprehension tasks. 
Paltenghi et al.~\cite{paltenghi2022extracting} found that CodeGen's~\cite{nijkamp2022codegen} self-attention layers attend to similar parts of code compared to human programmers' visual fixations when answering 
comprehension
questions about code. Similarly, Huber et al.~\cite{huber2023look} discovered overlaps in attention patterns between neural models and humans when repairing buggy programs.
Notably, Paltenghi and Prasdel~\cite{paltenghi2021thinking} compared language models' self-attention weights and humans' visual attention during code summarization. They found that model attention, measured by self-attention weights, does not align well with human attention. However, this work is limited by investigating only small CNN and transformer models. Most importantly, all aforementioned studies used white-box approaches towards interpretability of open-source models, limiting applicability to state-of-the-art proprietary models. 

Recently, Kou et al.~\cite{kou2023model} utilized both white-box and black-box perturbation-based approaches to measure LLMs' focus in code generation tasks, and discovered a consistent misalignment with humans' attention.
In general, these works have demonstrated that whether human and machine attention align depends heavily on the methods employed to approximate machine focus, as well as the specific code comprehension task examined. In this paper, we build upon former works by examining whether human attention correlates with feature attribution in language models, measured by a black-box perturbation-based approach, in code summarization.

\section{Experimental Design}
\subsection{Measuring Human Visual Focus} 
We used eye-tracking data measuring human attention from a controlled human study with 27 programmers. 
The study obtained IRB approval, and asked participants to read Java methods and write accompanying summaries~\cite{bansal2023modeling}. Each participant summarized 24--25 Java methods from the FunCom dataset~\cite{leclair2019recommendations}, yielding 671 trials of eye-tracking data in total. 
Considering data quality, two authors with five and eight years of Java experience cooperatively removed participant data associated with five summaries that did not demonstrate an understanding of the Java code. 

In this work, we sought to measure where humans and language models focus on code as they summarize it.
We first used the \texttt{srcML} parser to convert each Java method into its corresponding Abstract Syntax Tree (AST) representation \cite{collard2013srcml}. 
The AST  provides structural context for each token literal (i.e., `Hello World' $\xrightarrow{}$ String Literal). 
With the gaze coordinates collected from the eye-tracker \cite{Tobii_2023}, we measured humans' focus on each AST token.
Typically, researchers use \textit{fixations} to quantify human visual focus \cite{sharafi2015eye}. A fixation is defined as a spatially stable eye-movement lasting 100--300ms. Most cognitive processing occurs during fixations \cite{sharafi2015eye}, so researchers consider their frequency and duration in making inferences about human cognition. 
In our analyses, we computed the average count and duration of programmers' fixations on each AST token. Consequently, for each Java method, we obtained two visual focus vectors with lengths equal to the number of AST tokens, respectively, which represent fixation counts and durations on each token\footnote{Our analyses do not include brackets or semi-colons, or other such syntactic elements.}. 









\subsection{Measuring Model Focus}
As mentioned in Section~\ref{related:attention}, we choose SHAP's official, default implementation of the TeacherForcing method to measure feature attribution in language models, treating each as a black-box function. 
For each language model, we pass in each of the 68 Java methods (also read by human programmers) as input, along with necessary prompting for the model to output summaries of source code.
For each Java method passed into each language model, we let $i$ denote an input token (in code) and $o$ denote an output token (in summary). For each ($i$, $o$) pair, SHAP produces an importance score, denoted $v_{(i, o)}$, signifying how much $i$'s presence or absence alters the presence of $o$. Then, the importance score of each input token\footnote{We use the absolute value by choice, without which experiments show similar results.}, $v_i$, is calculated such that $v_i = \sum_{o} |v_{(i,o)}|$. Note that now $v_i$ is associated with a language model token, and each AST token may consist of several language model tokens. Thus, for each AST token, we calculate its score $\frac{1}{n}\sum_{j=1}^n v_j$, where $v_1, \cdots, v_n$ are scores of language model tokens constituting the AST token. Consequently, for each language model on each Java method, we obtain a focus vector (with a length equal to the number of AST tokens) representing how influential each AST token is to the model.

In total, we investigated the model focus of six different models: GPT4, GPT-few-shot, GPT3.5, StarCoder, Code Llama, and NCS. Here, GPT-few-shot is a GPT3.5 model, but in an attempt for the model to produce code summaries more similar to those of humans, we used few-shot prompting to instruct the model to provide summaries similar to two randomly selected human-written summaries. The other five state-of-the-art LLMs are introduced in Section~\ref{related:model} and implemented with their default parameters.

\subsection{Comparing Human and Model Foci}

For brevity, we refer to two human visual focus measurements (i.e., fixation duration and count) and six language models as eight "focus sources." For each source, we obtained 68 focus vectors, each corresponding to a Java method. 
These vectors were normalized to sum to $1$, and reflect how important each AST token is for the human/model.
We answer these research questions:  

\squishlist
    \item RQ1: Is there a general correlation between human and machine focus patterns for code summarization?
    \item RQ2: Do the code summaries increase in quality when machine focus becomes more aligned with that of humans?
\squishend

\subsubsection{RQ1} We assess the correlation between human and machine foci across the 68 Java methods.
Specifically, for each pair of focus sources,  we iterate through each Java method and calculate the Spearman's rank coefficient ($\rho$)~\cite{spearman1961proof} between the two sources' 
vectors for that method.
Then, for each pair of focus sources, we report: (1) The mean and standard deviation of Spearman's $\rho$ across all Java methods where correlation is statistically significant ($p\leq 0.05$), and
(2) the proportion of Java methods demonstrating a statistically significant correlation ($p\leq 0.05$).

In addition, we group all AST tokens into 18 
semantic categories (e.g., method call, operator, etc.) and investigate how much humans\footnote{We use fixation durations to represent human focus.
We empirically verify that using fixation count yields similar results.} and language models focus on each semantic category.
The focus score assigned to each semantic category is the sum of the focus scores assigned to each AST token belonging to that semantic category. 
To counter biases where certain semantic categories contain more AST tokens or appear more frequently, we report the relative difference between machine and human foci for each semantic category.
That is, we average the six language models' focus scores per category and report $|\frac{foc_{machine}- foc_{human}}{foc_{human}}|$.

\subsubsection{RQ2} Here, a human expert provides quality ratings for summaries generated by each language model for every Java method using four criteria: accuracy, completeness, conciseness, and readability.  Next, we calculate the Spearman's $\rho$ between each language model's focus vector and the human fixation duration vector across all Java methods where correlation is significant ($p\leq 0.05$). We then append all such statistically significant $\rho$'s to form a vector, denoted $v_{cor}$, to represent the degrees of alignment between machine and human foci across the Java methods investigated\footnote{Note that $v_{cor}$ contains Spearman's $\rho$'s obtained from all six language models. We empirically verify that conducting the analogous analysis for each language model separately yields a similar result.}. 

Subsequently, we determine whether this alignment is correlated with the rated quality of summaries. Specifically, we construct four other vectors, $\{v_{acc}, v_{com}, v_{con}, v_{rea}\}$, containing the accuracy, completeness, conciseness, and readability scores respectively. At each index $i$, $\{v_{cor}[i], v_{acc}[i], v_{com}[i], v_{con}[i], v_{rea}[i]\}$ are respectively the Spearman's $\rho$, summary accuracy, completeness, conciseness, and readability of the same language model applied on the same Java method. We then measure and report the Spearman's rank correlation between $v_{cor}$ and $v_i$, where $v_i\in\{v_{acc}, v_{com}, v_{con}, v_{rea}\}$.

\section{Results}\label{sec:results}

\subsection{RQ1: General Correlation}

As shown in Table~\ref{tab:pair_wise}, there is a general lack of correlation between human and machine foci. We highlight that the means and standard deviations in Table~\ref{tab:pair_wise} are only calculated from Java methods where the Spearman's $\rho$ is statistically significant (with $p\leq 0.05$). In practice, between any pair of human-LLM focus sources, at most $22\%$ of the 68 Java methods yield a Spearman's $\rho$ with $p \leq 0.05$. As a baseline, the Spearman's $\rho$ has $p\leq 0.05$ for all Java methods between human duration and fixation focus vectors, and for $85\%$ of Java methods between any two language model's focus vectors. This implies that any existing correlation between human and machine foci is not widespread across the Java methods studied.

Furthermore, among those Java methods where the correlation is statistically significant, the mean Spearman's $\rho$ is small with a large standard deviation. In fact, for most such methods where a model and human show significant correlation in focus, the Spearman's $\rho$ is often either around $0.5$ or $-0.5$, but rarely in between.
This suggests the relationship between human and machine foci varies significantly depending on the specific Java method.

Interestingly, although few-shot-alignment in GPT-few-shot renders the model's generated summaries more similar to those of humans, this does not lead to higher correlations between model and human foci. In addition, feature attribution in all language models is moderately or strongly positively correlated with each other on a majority of Java methods, which intuitively makes sense since all six models are based on the Transformer architecture.

\begin{figure}[tbp]
\centering
\includegraphics[width=0.9\linewidth]{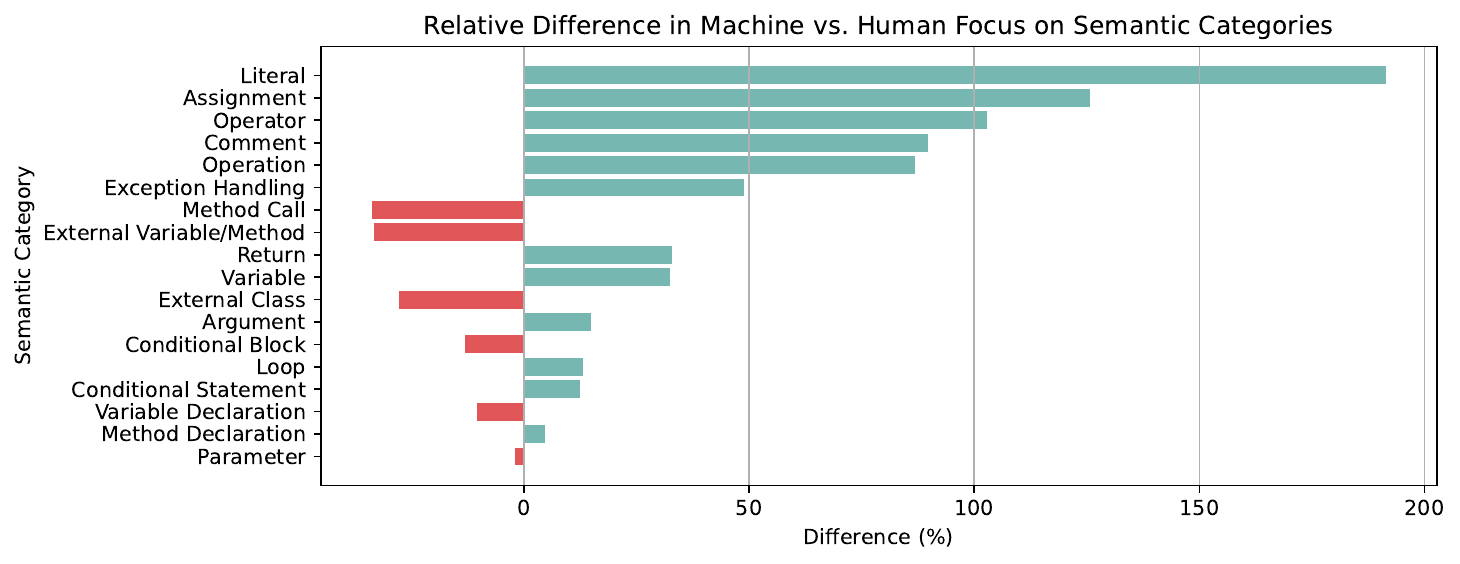}
\caption{How much more/less do language models focus on each semantic category compared to humans? }
\label{fig:semantic_category}
\vspace{-0.2in}
\end{figure}

We also investigate how language models' focus on each semantic category differs from that of humans. As shown in Figure~\ref{fig:semantic_category}, language models' generation of code summaries seems to be more reliant on comments, return values, and specific statements such as literals and assignments, and less reliant on method calls and variables/methods not defined explicitly within the Java method.

\noindent\textbf{Discussion Point 1:} We find no evidence that feature attribution in language models is correlated with programmers' visual focus. Several possible interpretations can be inferred: 
(1) Alternative methods may be needed to assess feature influence in black-box language models for code summarization, aiming for better alignment with human attention.
(2) Access to the internal workings of proprietary models might become critical if white-box models offer more human-aligned insights into explainable language models for code~\cite{paltenghi2022extracting}.
(3) It is possible that language models and humans reason about code differently when summarizing source code.


\begin{table}[tbp]
\addtolength{\tabcolsep}{-4.5pt}
\caption{Pair-wise correlation among focus sources; ``Duration'' and ``Count'' are human visual focus. 
Each cell shows the means and standard deviations of Spearman's $\rho$ for all Java methods showing significant correlation ($p \leq 0.05$).}
\label{tab:pair_wise}
\fontsize{6pt}{6pt}\selectfont
\begin{tabularx}{\columnwidth}{l|rrrrrrrr}
\toprule
  & \multicolumn{1}{c}{Duration} 
  & \multicolumn{1}{c}{Count} 
  & \multicolumn{1}{c}{GPT4}
  & \multicolumn{1}{c}{GPT-few}
  & \multicolumn{1}{c}{GPT3.5}
  & \multicolumn{1}{c}{StarCoder}
  & \multicolumn{1}{c}{CodeLlama} 
  & \multicolumn{1}{c}{NCS} \\
\midrule
\rowcolor{gray!20}

Duration     & 1.00$\pm$0.00     & 0.88$\pm$0.06     & -0.11$\pm$0.41    & -0.13$\pm$0.42    & -0.09$\pm$0.52    & -0.18$\pm$0.48    & -0.18$\pm$0.42    & -0.24$\pm$0.40 \\
\rowcolor{gray!20}
Count        & \multicolumn{1}{c}{---}               & 1.00$\pm$0.00     & 0.01$\pm$0.45     & -0.24$\pm$0.33    & -0.10$\pm$0.48    & -0.31$\pm$0.29    & -0.13$\pm$0.43    & -0.33$\pm$0.33 \\
GPT4         & \multicolumn{1}{c}{---}               & \multicolumn{1}{c}{---}               & 1.00$\pm$0.00     & 0.68$\pm$0.12     & 0.76$\pm$0.12     & 0.67$\pm$0.14     & 0.67$\pm$0.14     & 0.55$\pm$0.13 \\
GPT-few      & \multicolumn{1}{c}{---}               & \multicolumn{1}{c}{---}               & \multicolumn{1}{c}{---}               & 1.00$\pm$0.00     & 0.72$\pm$0.12     & 0.62$\pm$0.15     & 0.64$\pm$0.15     & 0.55$\pm$0.13 \\
GPT3.5       & \multicolumn{1}{c}{---}               & \multicolumn{1}{c}{---}               & \multicolumn{1}{c}{---}               & \multicolumn{1}{c}{---}               & 1.00$\pm$0.00     & 0.65$\pm$0.16     & 0.67$\pm$0.15     & 0.58$\pm$0.13 \\
StarCoder    & \multicolumn{1}{c}{---}               & \multicolumn{1}{c}{---}               & \multicolumn{1}{c}{---}               & \multicolumn{1}{c}{---}               & \multicolumn{1}{c}{---}               & 1.00$\pm$0.00     & 0.66$\pm$0.15     & 0.59$\pm$0.11 \\
Code Llama   & \multicolumn{1}{c}{---}               & \multicolumn{1}{c}{---}               & \multicolumn{1}{c}{---}               & \multicolumn{1}{c}{---}               & \multicolumn{1}{c}{---}               & \multicolumn{1}{c}{---}               & 1.00$\pm$0.00     & 0.56$\pm$0.14 \\
NCS          & \multicolumn{1}{c}{---}               & \multicolumn{1}{c}{---}               & \multicolumn{1}{c}{---}               & \multicolumn{1}{c}{---}               & \multicolumn{1}{c}{---}               & \multicolumn{1}{c}{---}               & \multicolumn{1}{c}{---}                 & 1.00$\pm$0.00 \\
\bottomrule
\end{tabularx}
\end{table}

\begin{figure}[tb]
\centering
\includegraphics[width=0.9\linewidth]{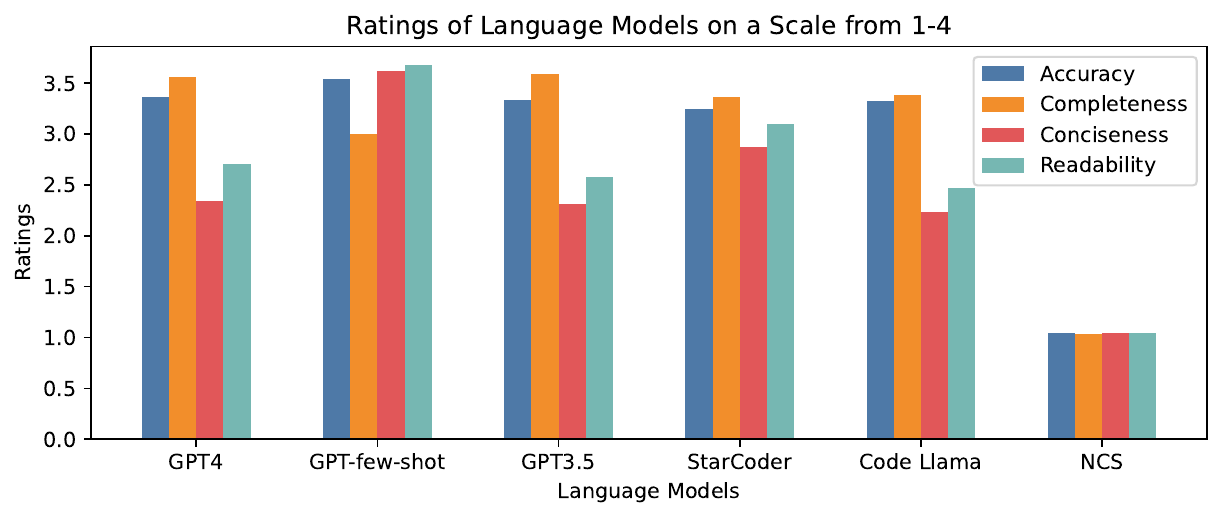}
\vspace{-16pt}
\caption{Average ratings of model-generated summaries.\vspace{-12pt}}
\label{fig:model_ratings}

\end{figure}

\begin{table}[b]
\vspace{-12pt}
  \caption{Correlation between human-machine focus alignment and summary quality (assessed by four metrics).\vspace{-6pt}}
  \label{tab:metrics}
  \scalebox{0.82}{
  \begin{tabular}{rrrrr}
    \toprule
    &  \multicolumn{1}{c}{Accuracy} 
    & \multicolumn{1}{c}{Completeness}
    & \multicolumn{1}{c}{Conciseness} 
    & \multicolumn{1}{c}{Readability} \\
    \midrule
    Spearman's $\rho$ &
   -0.1279 &  
   0.1309 &   
   0.0194 &   
   -0.0717   \\

   $p$-value &
   0.3862 &
    0.3753 & 
    0.8960 &
    0.6280 \\

  \bottomrule
\end{tabular}}
\end{table}

\subsection{RQ2: Summary Qualities}

There is also a lack of correlation between the quality of summaries generated by language models and how well their focus on code aligns with humans'. The large p-values in Table~\ref{fig:semantic_category} suggest that, regardless of which metric is used to assess summary quality, there is a lack of statistically significant correlation between the quality of a model-generated summary on a Java method and how well the model's focus aligns with that of humans on that Java method. Furthermore, Figure~\ref{fig:model_ratings} shows that NCS produces worse summaries than the other five models. Although Table~\ref{tab:pair_wise} seems to suggest that NCS's focus is more negatively aligned with human attention, we find no statistically significant metrics supporting such a claim, partially due to the small sample size of Java methods yielding statistically significant Spearman's $\rho$.

In general, Table~\ref{fig:semantic_category} suggests that feature attribution in NCS is still moderately positively aligned with that in other language models on a majority of Java methods. This indicates the likelihood that aspects other than feature attribution are more indicative of and critical to a language model's performance in code summarization.

\noindent\textbf{Discussion Point 2:} With a substantial body of work in NLP showing that aligning neural models with human visual patterns can lead to performance improvement \cite{sood2020improving, zhang2019using,klerke2016improving, barrett2018sequence}, we contain our conclusion to the SHAP measure of feature attribution and the human attention as measured in an eye-tracking experiment.  The link between human attention and feature attribution to machine models is a subject of intense scientific investigation.  We contribute to the debate with this finding that SHAP did not correlate with human eye attention in the measures or models we studied.

\section{Conclusion}

In this paper, we use a state-of-the-art, black-box, perturbation-based method to assess feature attribution in language models on code summarization tasks. We then compare the model-determined important AST tokens with those identified by human visual focus, as measured through eye-tracking. The results suggest that using SHAP to measure feature attribution does not provide explainability of language models through establishing correlations between machine and human foci. Generally, our work can be interpreted in two ways. First, feature attribution measured by SHAP may not be the best way to interpret a language model's focus during code summarization as it fails to establish similarities with human focus. Alternatively, it may be the case that machines reason about code differently from humans when tasked to summarize source code.

\section{Acknowledgement}
This research was supported by NSF CCF-2211429, NSF CCF-2211428, NSF CCF-2100035 and NSF SaTC-2312057.

\bibliographystyle{acm}
\bibliography{./main.bib}
\end{document}